\documentclass[12pt,draftclsnofoot,onecolumn]{IEEEtran}

\usepackage{lineno}
\usepackage{amsfonts,amsmath,amssymb}
\usepackage{dsfont}
\usepackage{cite}
\usepackage{graphicx}
\usepackage{url}
\usepackage{float}
\usepackage{stfloats}
\usepackage{enumerate}
\usepackage{tabularx}
\usepackage{multirow}
\usepackage{hhline}
\usepackage{booktabs}
\usepackage{makecell}
\usepackage{bm}
\usepackage{algorithm}
\usepackage{algpseudocode}
\usepackage{graphics}
\usepackage{epsfig}

\allowdisplaybreaks

\begin{document}
\title{Cybertwin: An Origin of Next Generation Network Architecture}
\author{Quan~Yu,~\IEEEmembership{Senior Member,~IEEE}, Jing Ren, Yinjin Fu, Ying Li, and~Wei~Zhang,~\IEEEmembership{Fellow,~IEEE}
\thanks{
Q. Yu is with Peng Cheng Laboratory, Shenzhen, China (email: yuq@pcl.ac.cn). 
J. Ren is with School of Information and Communication Engineering, University of Electronic Science and Technology of China, Chengdu, China, (email: 
renjing@uestc.edu.cn).
Y. Fu is  with Peng Cheng Laboratory, Shenzhen, China (email: fuyj@pcl.ac.cn). 
Y. Li is  with Peng Cheng Laboratory, Shenzhen, China (email:  liy02@pcl.ac.cn). }
\thanks{
Corresponding author:  W. Zhang is with Peng Cheng Laboratory, Shenzhen, China, and School of Electrical Engineering \& Telecommunications, University of New South Wales, Sydney, Australia (email:  w.zhang@unsw.edu.au).}
}

\maketitle

\begin{abstract}
 With fast development of Internet of Everything (IoE) and its applications,  the ever increasing mobile internet traffic and services bring unprecedented challenges including scalability, mobility, availability,  and security which cannot be addressed by the current clean-slate network architecture.   
 In this paper, a cybertwin based next generation network architecture is proposed to accomodate the evolution from end-to-end connection to cloud-to-end connection in the future network. As a digital representation of human or things in the virtual cyberspace,  cybertwin serves in multiple capacities, such as communications assistant, network data logger, and digital asset owner.  The new and unique characteristics of the cybertwin make the proposed network to be flexible, scalable, reliable, and secure. Further, we advocate a new cloud network operating system which can work in a distributed way through a real-time multi-agent trading platform to allocate 3C (computing, caching, communications) resources. We also propose cloud operator, a new operator that can provide and manage the resources to the end users and offer location and authentication services for human and things in the cyberspace.  Some promising and open research topics are  discussed to envision the challenges and opportunities of the cybertwin in the future network architecture.
   
\end{abstract}

\begin{IEEEkeywords}
Cybertwin, Cloud Network, Cloud Operator,  Network Architecture, Internet of Everything.
\end{IEEEkeywords}

\section*{Introduction}
\label{sec.Introduction}
  
Next generation networks will face the demand of billions of people and hundreds of billions of devices to be connected. Accordingly, the future network architecture is expected to accommodate the explosively increasing mobile internet traffic and various services and applications through heterogeneous networks.    
Internet of Everything (IoE) has been considered as the future of Internet and could achieve intelligent connections of human, process, data and things \cite{IoE}, with the help of fifth generation mobile communication (5G) and artificial intelligence (AI) technology  to make networked connections more relevant and valuable than ever before.   
  In comparison with the traditional end-to-end connection, a revolutionary feature of the network architecture in IoE is to support ubiquitous data collection, aggregation, fusion, processing, distribution and service \cite{IoEvalue}. The disruptive change raises many issues and challenges for the network architecture design of IoE, e.g., scalability, mobility, security and availability, etc.

In the current Internet architecture, the IP protocol uses IP address   for both  identity and  locator of the device attached to the network, so it cannot address the dramatically increasing demands of mobile devices and services  thereby leading to the scalability problem.  
In addition, the network trustworthy depends on not only the security of the end-to-end physical connections, but also the trusted user who accesses the network. However, in the current Internet architecture the anonymous access compromises the network trustworthy which results in the security issue.  
 Furthermore, in the current network architecture  it is difficult to coordinate the network resources among multiple network service providers to offer the personalized quality of service (QoS) guarantee and availability during a single communication process which leads to the availability issue. 
 All these challenging issues in the current network architecture are critical and severely impeding the fast growing development of mobile traffic and services.  
 
 To address the challenges,  
 Information-centric networking (ICN) architectures were proposed, such as Named Data Networking \protect \cite{NDN}, Content-Centric Networking \protect \cite{CCN}, DONA \protect \cite{DONA}, but these designs are incompatible with the existing IP network infrastructure and cannot well support mobility due to the unchanged end-to-end principle. 
 {MobilityFirst \protect \cite{MobilityFirst} is motivated by accessing from mobile platforms, but it neglects  the availability and security management. The eXpressive Internet Architecture (XIA) project\protect \cite{XIA} aims to  improve both the evolvability and trustworthiness of the Internet by using a set of control and management protocols that deliver trustworthy network service to users, but it does not consider independent availability and cost-efficient scalable network management.   
 ChoiceNet\protect \cite{Choicenet}  introduces a new economy plane to support advertisement of choices to users, but it does not provide real-time  trading for short-term contracts.
 
Recently, cloud-computing-centric network architectures were provided to achieve network resource sharing, to handle big data explosion from IoE devices, and to simplify management tasks \cite{pan2011survey}, such as Nebula \cite{nebula}, CloNe \cite{CloNe},  and Cloud Integrated Network \cite{weldon2016future}.     NEBULA \cite{nebula} provides resilient networking services (including dependability, security, flexibility and extensibility) for the future Internet architecture using ultra-reliable routers, an extensible control plane and use of multiple paths upon which arbitrary policies may be enforced. However, its scalability and performance are still limited due to the ignorance of processing capability at network edge. Further, it does not consider the issue of mobility support to accommodate the increasing demands of mobile devices and services.  
The CloNe architecture \cite{CloNe}  introduces a cloud networking architecture in a multi-administrative domain scenario, where network and data centre domains   interact to provide a dynamic elastic network connection service to cloud customers. Further, it deploys computing and storage resources distributed in the network to allow for better end-user experience and to lower the dependency on network capacity. However, it does not address the mobility services in the cloud networking and its simple access control policy selection cannot satisfy high-level security goals. 
The cloud integrated network (CIN) architecture \cite{weldon2016future} embeds the edge clouds into future network to provide the optimum performance and economics for both virtualized networking functions and other performance critical web services. 
It introduces a network operating system (OS) to abstract all network resources and manage the utilization to create exactly the needed capacity from the source to the destination.   
But it does not solve the challenges on mobility and availability in cloud networking. 
 
 Clearly there are many challenges  in providing scalability, security, mobility  and network security for future network architecture designs.    
In this article, we propose a cybetwin based cloud-centric network architecture for future generation networks. By introducing cybertwin, which serve as communications assistant, network data logger, and digital asset owner of human and things in the IoE, we design a new network architecture to address  scalability, security, mobility, and availability of future generation networks. 
  
 \section*{Cybertwin based Network Architecture}
\label{sec.arch}
 
\begin{figure}[!t]
\centering
\includegraphics[width = 5in]{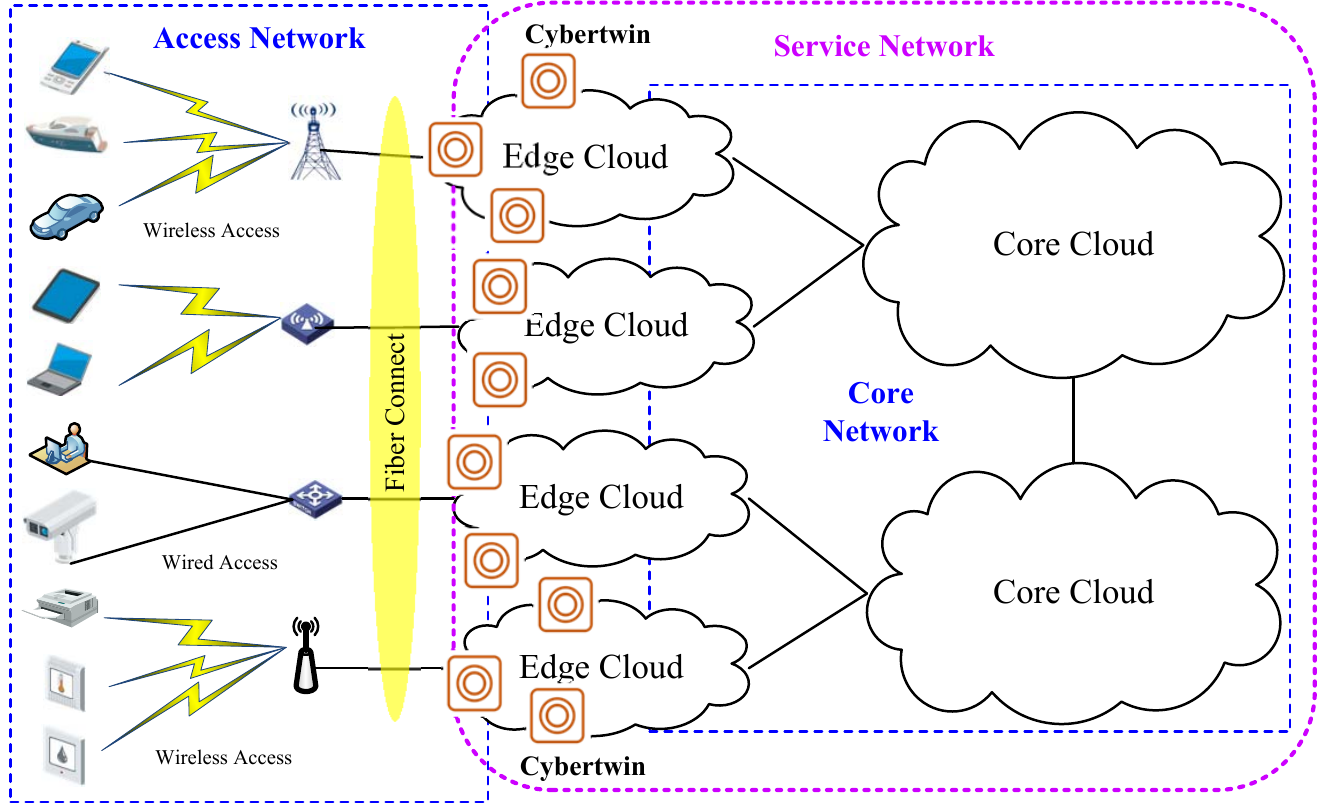}
\caption{Cybertwin based Cloud-Centric Network Architecture }
\label{fig:arch}
\end{figure}

Fig. \ref{fig:arch} shows our proposed cybertwin-based next generation network architecture. In this new architecture, there are four key components, i.e., Core Cloud, Edge Cloud, Cybertwin and the Ends, the functions of which are summarized as follows.

\begin{itemize}
	\item \textbf{Core Cloud} is  fully-connected with each other to build the core networks through high-speed optical links. Unlike the existing cloud networks, e.g., Amazon AWS, which only provide cloud computing service as a  particular application services, these core clouds in our proposed architecture provide computing, caching and communications resources for the ends as a network infrastructure service.
	\item \textbf{Edge Cloud}  resides between the core cloud and the ends. Although the resource it can provide is much less than the core cloud, it can respond faster to the ends' request due to its proximity to them. Thus, the edge clouds can help the core clouds in providing a very high quality of network services.
	\item \textbf{Cybertwin} is the most critical network function provided by the proposed network architecture. It is a digital representation of human or things  in the virtual cyberspace and is located at the edge cloud. It serves in multiple capacities, i.e., communication assistant, network data logger, and digital asset owner.  It provides a new cybertwin based communication model to replace the end-to-end communication model.  Details of cybertwin will be expounded in the next subsection. 
	\item \textbf{The Ends} refer to human and things in the network. They are consumers of the network services and are connected to the network via various access methods.   In the proposed new network architecture, the ends no longer need to establish a connection with a specified server following the traditional end-to-end communication model. Instead, the ends acquire services directly from the network through the cybertwin. 
\end{itemize}

To support the above four components, three kinds of networks, i.e., access network, core network, and service network, are employed  in the proposed network architecture design.  
\begin{itemize}
	\item \textbf{Access Network} is responsible for connecting the ends and edge clouds. Both wired and wireless access techniques can be used. 
	\item \textbf{Core Network} is in charge of establishing the connections between edge clouds and core clouds, as well as the connections among all core clouds.  
	\item \textbf{Service Network}  is a logical network for application services which are deployed on top of the cloud network to provide services to the ends.
\end{itemize}

These three kinds of networks are operated by telecommunications operators, cloud operators and application service providers, respectively. 
\begin{itemize}
	\item \textbf{Telecommunications operator} offers high-speed communication pipes between the ends and clouds in the access networks and among different clouds in the core networks. It can sale or lent its networking resources to the cloud operators. 
	\item \textbf{Cloud operator} is a new operator who provides information infrastructure services, not the general information services.  Some of its key features are explained below. 
	\begin{itemize}
	\item It provides computing, caching and communications resources to the ends by replacing or compressing the existing core networks, cloud computing service providers, cloud storage service providers (e.g., AWS, Azure, Baidu, Al, etc.), and CDN service provider, etc. 
	\item It provides a brand new operating environment for cybertwin.
	\item It provides function of  authentication  for human and things  via biological features and online behaviour, etc.
	\item It provides various trust levels for human and things. 
		\item It provides the real-time location service to human and things. 
	\end{itemize}

	\item \textbf{Application service provider} delegates their application and content services to the cloud operator which deploy these services in the edge and core cloud.  Thus, it can provide high quality of service and reduce its cost as there is no need to implement a dedicated server. 
	
\end{itemize}

\begin{figure}[!t]
\centering
\includegraphics[width = 5in]{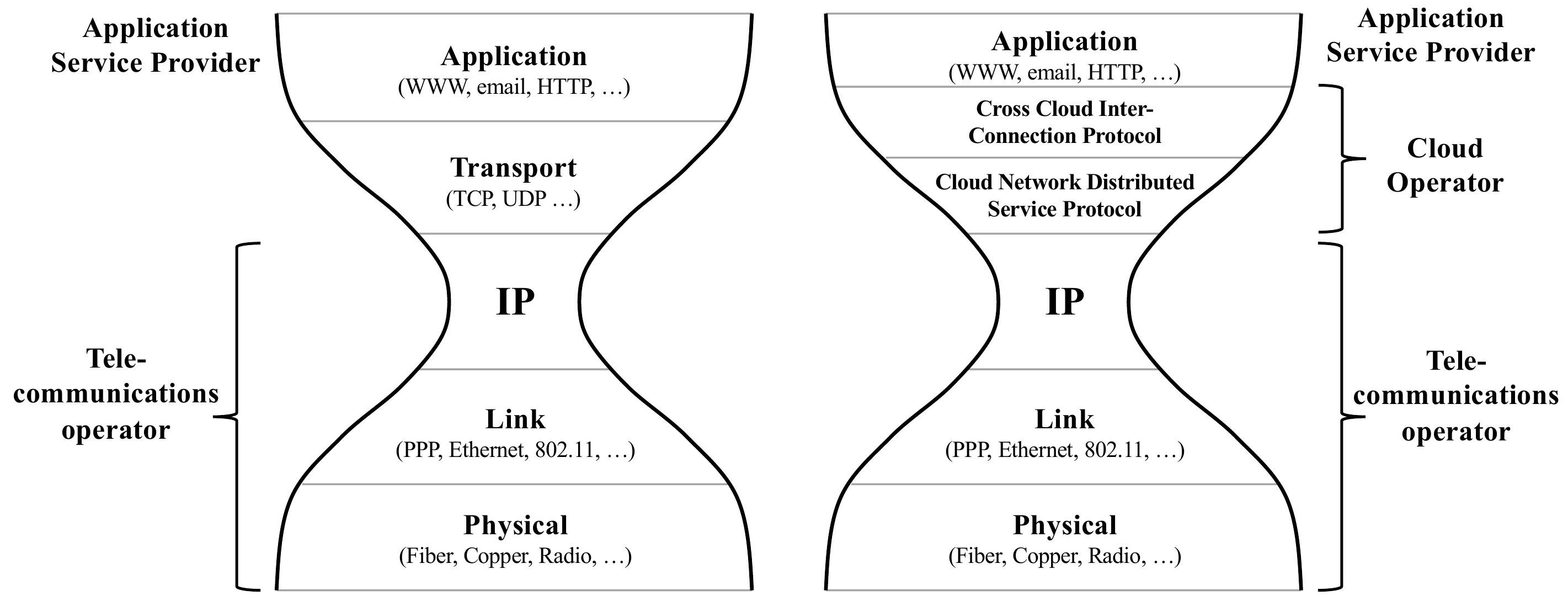}
\caption{The main building blocks of sandglass model for new network architecture}
\label{fig:protocol}
\end{figure}

Compared with the traditional Internet, we keep the IP layer as the ``thin waist'' of the stack but allow the evolutions of other layers (as shown in Fig. \ref{fig:protocol}). 
This means we inherit the connectless packet switch mechanism and allow the incremental updates of legacy network infrastracture.   
The transport protocol is replaced by the cloud network distributed service protocol and the cross-cloud operator interconnection protocol. The cloud network distributed service protocol is designed to deal with the multi-dimensional resources including computing, caching, and communications resources of a cloud operator in a distributed way. Based on this protocol, a cloud network operating system can be built for each cloud operator. The cross-cloud interconnection protocol is used to provide network services through the cooperation of different cloud operators.

\subsection*{Cybertwin }

In the proposed network architecture design, Cybertwin is the key component to replace the traditional end-to-end communication model.    As shown in Fig. \ref{fig:cyberTwin}, it has three fundamental functions: communication assistant, logger and digital asset as listed as follows.

\begin{itemize}
	\item \textbf{As communication assistant}, Cybertwin is  the digital representation of human or things in the virtual world and also acts as the virtual network ID of human and things. As the assistant of human or things in the physical world, cybertwin can better achieve various QoE because it knows well the users' QoS and will pay the cost to negotiate with telecommunications operator and cloud operator for the resource. 
		Unlike establishing the end-to-end connections between the end and the server which provides the services, the cybertwin based communication model requires the ends to first connect to its cybertwin that will acquire the required service from the cloud network on behalf of the ends. The required service is provided by the edge cloud and core cloud jointly through the distributed cloud network operating system which pooling various  computing, caching, and communications resources. To provide high quality of service, cybertwin may request the service from multiple locations simultaneously.  	\item \textbf{As data logger}  In the proposed network architecture design, communication should be completed through cybertwin, it means that the cybertwin can obtain and log all data about the end’s behaviours in real-time fashion. Traditionally, these data are acquired and stored by a few companies and may be abused without the permission of the ends. However, cybertwin is the digital representation of the ends and does not belong to any particular company. 
	\item \textbf{As digital asset owner},  cybertwin can also convert these logged behaviour data into a digital asset by processing these data as a service and publishing the service to other entities including the application service providers and cloud operators. This service can break the monopoly in which only a few big companies gather these valuable data. Further, it can also bring economic value to the owners of the data and may give birth to a new business model C2X (customer-to-everything). 
 \end{itemize}

\begin{figure}[!t]
\centering
\includegraphics[width = 3.6in]{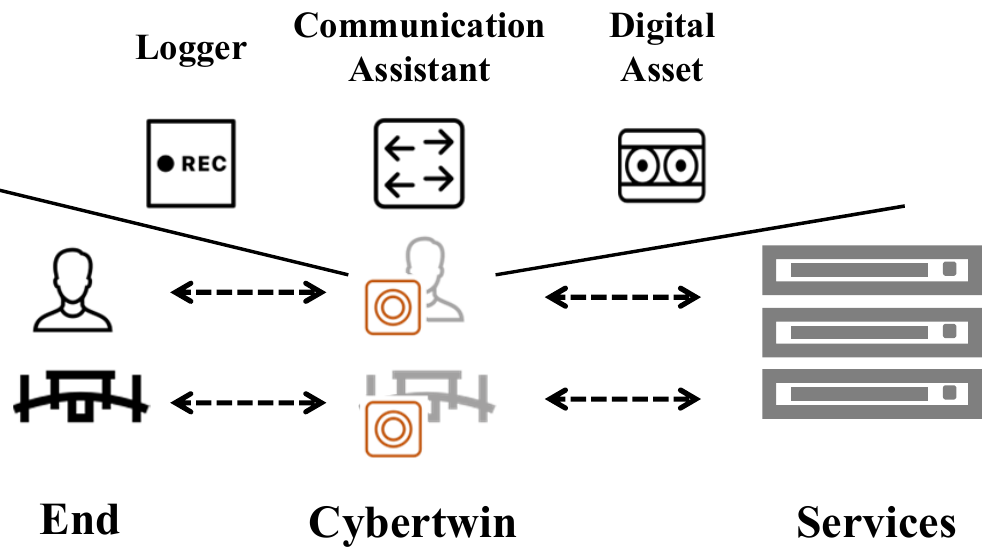}
\caption{Cybertwin based communication}
\label{fig:cyberTwin}
\end{figure}

One of the  benefits of using cybertwin is to support the locator/identifier separation. In the existing networks, the bottleneck problem to implement locator/identifier separation is the lack of an identity authentication mechanism. Without the authentication mechanism, the networks cannot know whether a user is the one who  declares.  
 As a communication assistant,  cybertwin has its natural ability to provide identity authentication.  

More benefits of cybertwin based communication model, including security, mobility, availability and scalability, will be discussed in details in next Section.

\subsection*{Cloud Network Operating System}
 
A massive number of users and large scale cloud-centric networks in the future generation networks pose great challenges in the allocation of  multi-dimension resources, i.e., computing, caching and communications resources. In the existing network architecture, the resources are scheduled and utilized  independently with each other, resulting in the inefficiency and latency problems. Further, the centralised management and allocation of the multi-dimension resources becomes infeasible considering a large number of users, various cloud networks, and multi-operators which hardly cooperate via a centralised mechanism. 

We advocate a new cloud network operating system which can work in a distributed way via establishing a real-time market driven  trading platform for multi-agents (i.e., the ends, telecommunications operators, applications service providers, and cloud operators.) The current blockchain mechanism can provide the function of de-centralisation but inherits the bad real-time problem. We propose to use the real-time contract signed by cloud operators to achieve real-time transactions among multi-agents. On this platform, the multi-dimensional resources will be priced based on the scarcity of these resources dynamically following some economic principles, e.g., traffic demand, importance and popularity of the services, etc. When some resources become scarce, the resource will be re-allocated according to the smart contract. Some other methods, e.g., big data analytics, machine learning, network billing and so on, are also employed to help for pricing and carrying out contract.

\begin{figure}[!t]
\centering
\includegraphics[width = 5in]{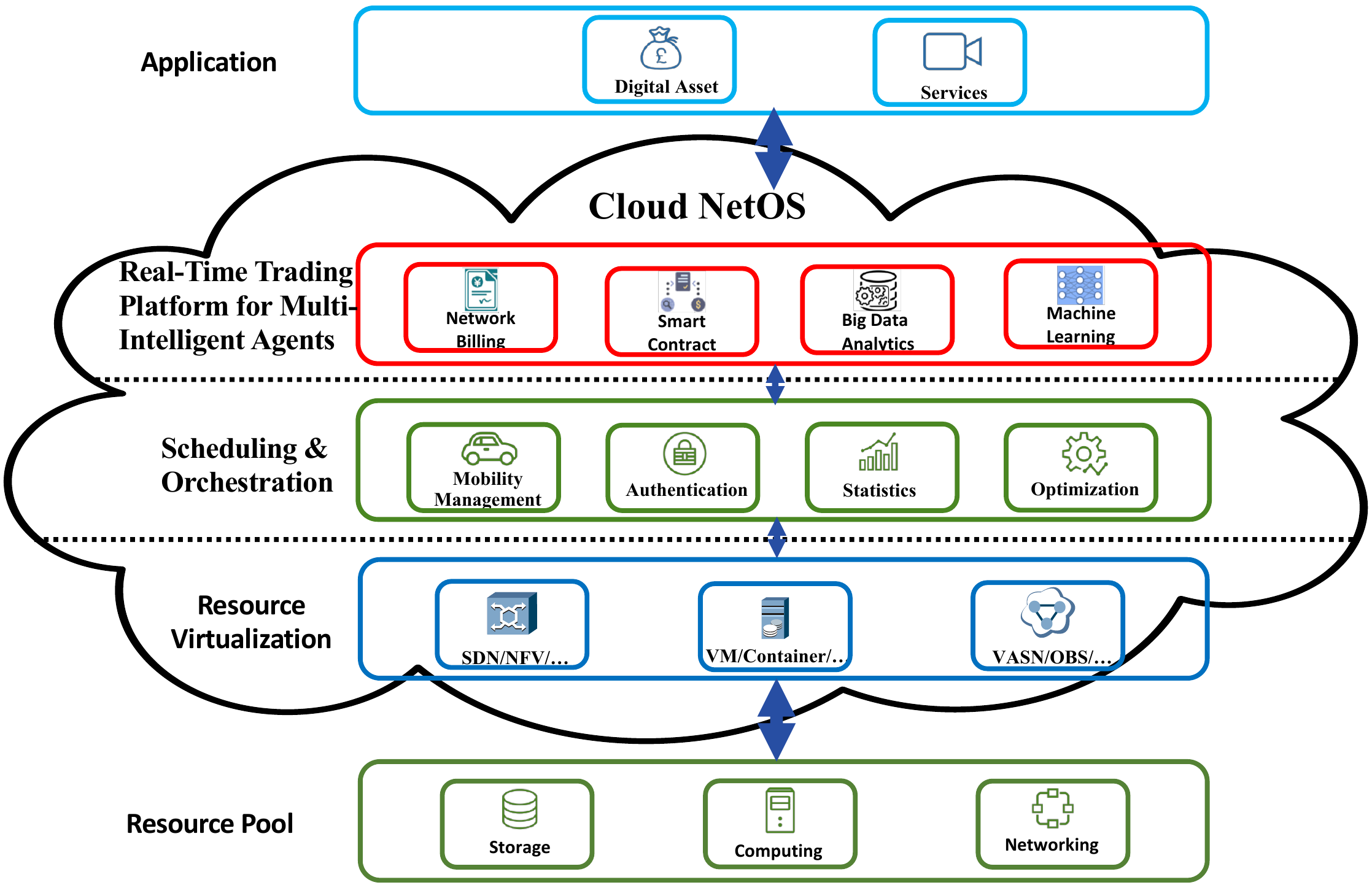}
\caption{Cloud Network Operating System}
\label{fig:cloudOS}
\end{figure}

To support this real-time trading platform, resource virtualization methods like SDN and NFV can be employed in a limited scope and used as the basis for the cloud network operating system that runs in a broader range in a distributed way, as shown in Fig. \ref{fig:cloudOS}. Some other technologies, including mobility management, authentication, statistics, and optimization, are applied for providing the capabilities of scheduling and orchestration, which use the virtualized resource to perform the contract signed through the real-time trading platform. As illustrated in Fig. \ref{fig:cloudOS}, application services and  digital assets data service can be performed on top of the cloud network operating system.

\section*{Features of Cybertwin based Next Generation Network Architecture Design}
\label{sec:Feature}

Our cybertwin based future network design has a number of new and powerful features in terms of  mobility, security, availability, and scalability, etc. We show these features next.

\subsection*{Mobility}
It was recently predicted in \cite{cisco2018cisco} that global mobile data traffic would grow nearly twice the growing speed of  fixed IP traffic. This fundamental shift from fixed to mobile traffic in Internet usage represents a unique and timely opportunity to take into account the special requirements of mobile devices and applications. While adding mobility to the current Internet architecture is still challenging, as the current Internet naming system is based on the host address, typically the IP address. In our network architecture design, 
we use the object identifers (ID) to name human, things and services, and use network address to indicate their locations. 
We also propose to build a mapping between object IDs and network address in cyberspace to support mobility. This mapping can be dynamically managed in the core clouds and edge clouds. The end devices gets the mapping from an edge cloud, and the edge cloud further obtains the mapping information from a core cloud. Then, we can build a cyber-physical-social systems (CPSS) indexing to exhibit the mobility of human or things and the location of network services.

The process of the primary communication and the mobility management with cybertwin is shown in Fig. \ref{fig:mobility} and listed in following steps. 

\begin{figure}[!t]
\centering
\includegraphics[width = 5in]{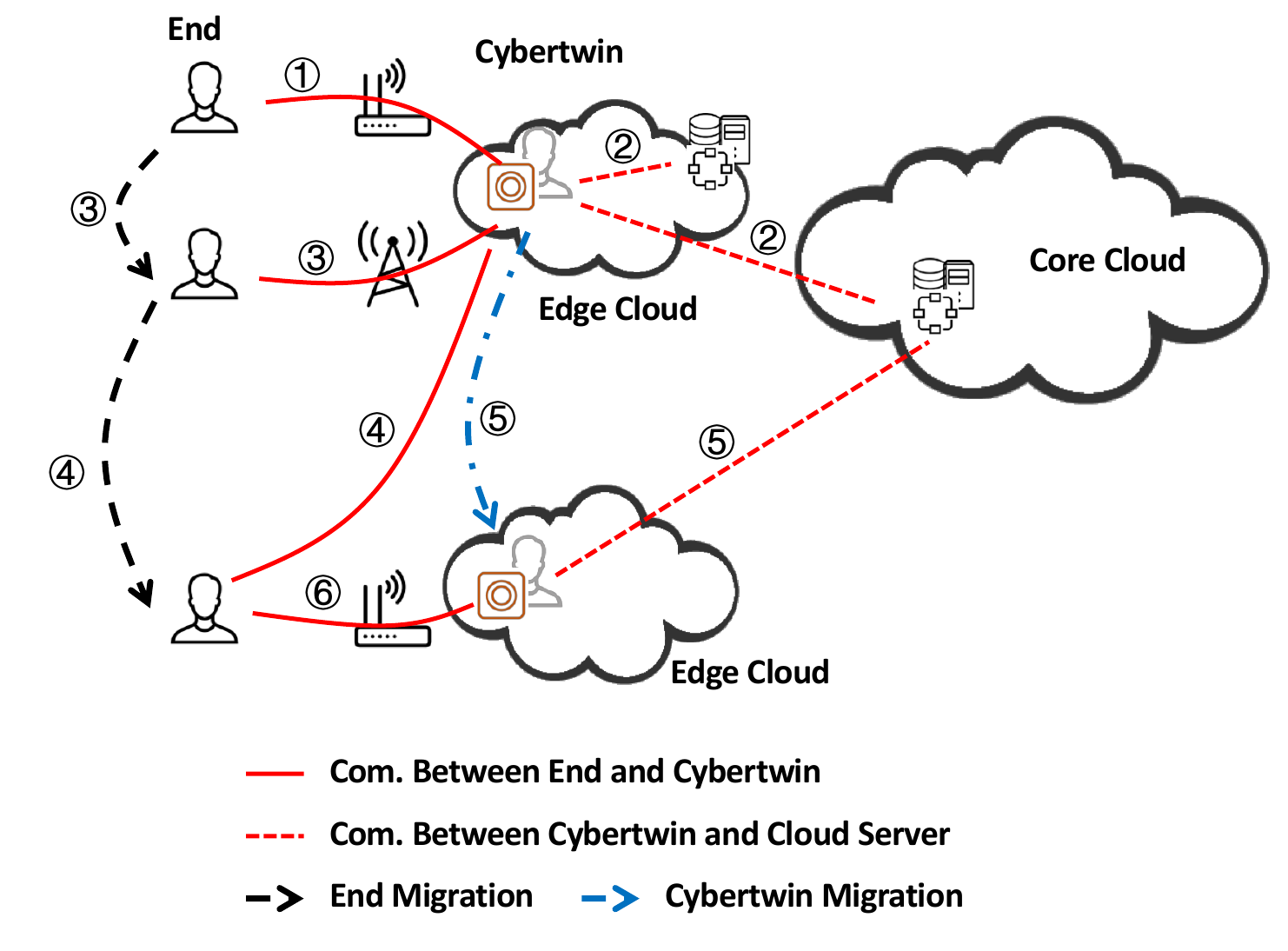}
\caption{The process of mobility management with cybertwin}
\label{fig:mobility}
\end{figure}

\begin{itemize}
\item[\textcircled{1}]  When an end requests a service, it first connects to its cybertwin hosting at the edge cloud.
\item[\textcircled{2}]  The cybertwin gets the collaborated service from the edge cloud and core cloud according to the scheduling of the cloud network operating system. Then, it passes on the required service data to the end.
\item[\textcircled{3}] When an end moves to a new location, the movement only requires the end to connect to its cybertwin through a new access point. When the end disconnects from the original access point, the cybertwin can cache the service data and continue to provide these data to the end when it connects to the cybertwin again. 
\item[\textcircled{4}] If the end moves further to a new position, it can connect to the original cybertwin, which may lead to a longer delay.
\item[\textcircled{5}] The cybertwin may also need to migrate to another edge cloud that is closer to the end, and establish a new connection to acquire the required service. 
\item[\textcircled{6}]  After the migration, the end will connect to the new cybertwin.  The end could still keep the connection with the original cybertwin so as to choose the best cybertwin.

\end{itemize}
 
The above procedure shows that  cybertwin can perform quick backup and real-time migration to support high availability and low latency and it is an always-on connection to serve the ends uninterruptedly.

\subsection*{Security}

Security includes both network security and privacy protection.  
Since anonymity is the core issue of Internet security, cybertwin can naturally and essentially provide security services in its design by exploiting behaviour tracking and authentication service, and further defending against DDoS attack and IP address hijacking.  
In our design, the locator and identifier separation scheme makes the network address becomes an internal address, which is opaque to the external network for safety. Furthermore, we combine the naming mechanism with an authentication mechanism to ensure there is no imposter. Different authentication methods can be employed for different objects. For example, we can identify human with its biological characteristics, and apply digital signature for thing or service authentication. A unified object identification can also help improving the network security with cyberspace real-name system. With the help of cybertwin, the behaviour of anyone or anything in cyberspace is audit-able and traceable. Furthermore, cybertwin can  obtain and log all data about the end’s behaviours in real-time fashion, thereby giving extra protection to the personal privacy.

\subsection*{Availability (QoS guarantee)}
Today’s Internet supports many critical applications that underpin the foundations of our modern society. However, the current state of availability of the Internet is far from being commensurate with its importance. As more critical processes are conducted as part of the IoE, the need for availability in IP networks increases. Any interruption to the transmission of data over networks negatively impacts these processes. Mission critical applications such as augmented reality require ultra-high reliability and ultra-low latency, while others like process automation are less demanding. Due to the quality of service (QoS) requirements are indeed application-dependent and various, IoE presents different QoS requirements from conventional homogeneous networks.

In the proposed cloud centric design, cybertwin can support always-online service when an end device accesses the edge cloud in access network. It can also provide the collaboration service for application or data content among the core clouds and edge clouds with high availability and low latency. A cloud network OS is provided to perform distributed collaboration decision with the economic principles and big data mining to meet the demand on intelligent adaption and load balance between the end needs and cloud resources. We provide a series of dynamic price adjustment rules based on the occupancy rate of resource in computing, caching and communicaitons, and define the corresponding economic price (e.g. traffic fee) for these network resources. With the help of cybertwin, we can provide the personalized QoS guarantee for various ends. 

In addition, the cross-cloud cooperative service can also be implemented among multiple cloud operators. It becomes possible to make trade among ends, application service providers, cloud operators and telecommunications operators using smart contracts based on blockchain technology. Furthermore, in multiple cloud operator scenario, a cross cloud operator interconnection protocol is needed with some principles in simplicity and openness, de-centralized autonomy and centralized cooperation to embody network territory and network sovereignty.

\subsection*{Scalability}
Future internet is expected to connect billions - and someday trillions - of objects and support information exchange among them. However, the scalability challenge of Internet stems from the conflict between the unprecedentedly high traffic surge and the exhaustion of IP address and the explosion of corresponding route table. Content delivery network (CDN) can partially alleviate the network traffic jam by duplicating and caching data locally, but cannot thoroughly solve the problem due to its high cost and inflexibility.   The network address translation (NAT) and IPv6 are the effective methods to mitigate the IPv4 address exhaustion, but these methods incur extra complexity. 
 
Our proposed cybertwin based cloud-centric  network architecture can address the scalability problem. Different from traditional cloud-network-end model, Cloud-centric architecture provides a cloud-centric direct communication model to shorten the network access path. Furthermore, our Cloud-centric design has a low complexity by pushing the intelligence, processing power and communication capabilities down to edge clouds or devices near the end, such as a node in local area network, an edge gateway. It can avoid long-distance access to the cloud and reduce the network access latency by responding to the user requests as near the end as possible.  
Hence, our Cloud-centric architecture can alleviate the route table expansion by improving the network access efficiency.

In the Cloud-centric architecture, we leverage the locator/identifier separation scheme to offer benefits on scalability using two independent namespaces for naming and addressing. It use an object ID with 128 bits IPv6 format or 48bits EUI-48 format to identify human, devices, services (data and apps) in the cyberspace. The network address, such as IPv4/ IPv6, becomes an internal address, which is opaque to the external network. We can build a mapping between object IDs and network address in cyberspace to link the ends with access point and save the IP addresses. Hence, the address exhausted problem can be alleviated by the mixed usage of object IDs and network address.

\section*{Conclusion and Future Work}
In this article, we have proposed cybertwin based next generation network architecture design. We have introduced new concepts cloud operators and cybertwin in the proposed network architecture. We have demonstrated the critical concepts of the architecture and provided four features to illustrate the value of cybertwin. To achieve a more safe and scalable architecture, some  open topics should be further explored.
\begin{itemize}
	\item \textbf{Identity authentication}: It  is critical to network security as it makes the network behaviour traceable. There are three types of objects including human, things, and services needed to be authenticated. Therefore, some efficient and trusted authentication methods or frameworks should be developed, such as biological characteristics based identification for human.
	\item \textbf{Digital asset management}: There are a vast number of digital asset hosting in a large number of cybetwins. It is essential to develop a digital asset management framework to standardise metadata model for the digital asset, to catalog the digital asset, and to perform permission of the digital assets' usage.
	\item \textbf{Cross-cloud operator resource scheduling}: It is a vital issue for the new future network design. A series of inter-operation principles shall be determined to guide the cooperative resource scheduling among different cloud operators. 
	\item \textbf{AI power based cybertwin}: Since cybertwin is the digital representation of the human or things in cyberspace, AI techniques can be applied into the deployment and collaboration among multiple cybertwins for high efficiency and QoS guarantee.
	\item \textbf{Smart contract based cloud resource coordination}:  Blockchain technology has emerged as a solution to consistency problems in peer to peer networks. Smart contract based coordination becomes an attractive mechanism for resource management in distributed systems of single-cloud or cross-cloud operator, without the need for a trusted (physical) third party.
	\item \textbf{Data privacy protection and security}: Cybertwin can obtain all data about the end behaviours, and convert these data into a digital asset by wrapping these data as a service and publishing the service to other entities. Hence, it becomes a critical challenge for the data management. We urgently need to formulate relevant laws and techniques to guide and protect the data privacy and security.
\end{itemize}

%\bibliographystyle{IEEEtranBST/IEEEtran}
%\bibliography{bib}

\end{document}